\documentstyle[11pt]{article}

\def\s{\stackrel}
\def\m#1{_{\mu_{#1}} }
\def\l#1{_{\lambda_{#1}} }
\def\lm#1#2{_{\lambda_{#1},\mu_{#2}} }
\def\ml#1#2{_{\mu_{#1},\lambda_{#2}} }
\def\dval#1#2{_{\lambda_{#1},\lambda_{#2}} }
\def\dvam#1#2{_{\mu_{#1},\mu_{#2}} }
\def\ep#1{^{\epsilon_{#1}} }
\def\barep#1{^{\bar{\epsilon}_{#1}} }
\def\k#1{_{\kappa_{#1}} }
\def\n#1{_{\nu_{#1}} }
\def\mk#1{{_{\mu_{#1}^{\kappa}}} }
\def\mn#1{{_{\mu_{#1}^{\nu}}} }
\def\lk#1{{_{\lambda_{#1}^{\kappa}}} }
\def\lnnew#1{{_{\lambda_{#1}^{\nu}}} }
\def\sig#1{_{\sigma(#1)} }

\begin{document}
\begin{titlepage}
\begin{center}
{\it P.N.Lebedev Institute preprint} \hfill FIAN/TD-20/93 \\
{\it I.E.Tamm Theory Department} \hfill December 1993

\vspace{0.1in}{\Large\bf Comultiplication in ABCD algebra \\
and scalar products of Bethe wave functions}\\[.4in]
{\large  A.Mikhailov}
\footnote{Supported in part by the RFFR Grant No.93-02-14365}\\
\bigskip {\it  The Moscow State University\\
Physical department\\
Moscow, Lenin hills, Russia\\
mika@td.fian.free.net}

\end{center}
\bigskip \bigskip

\begin{abstract}
The representation of scalar products of Bethe wave functions
in terms of the Dual Fields, proven by A.G.Izergin and V.E.Korepin in 1987,
plays an important role in the theory of completely integrable models.
The proof in \cite{Izergin87} and \cite{Korepin87} is based on
the explicit expression for the "senior" coefficient which was guessed
in \cite{Izergin87} and then proven to satisfy some recurrent relations,
which determine it unambiguously. In this paper we present an alternative
proof based on the direct computation.
\end{abstract}

\end{titlepage}

\section{Introduction.}

The Algebraic Bethe Ansatz~\cite{FaddeevTakhtajan,book} is the powerful  method
in the analysis of completely integrable quantum models. In the series of
papers
{}~\cite{Korepin82,Korepin84b,Korepin84a,Korepin87} various representations of
correlators in these models were found using this method. The answer turned
to be very interesting. Usually these correlators have some determinant
representations (see proof of Gaudin hypothesys in ~\cite{Korepin82}), and
in some "free fermionic point" these determinants are the
$\tau$-functions ~\cite{SW,DJM,Mor}
of classical nonlinear integrable equations ~\cite{IIKSdiff,IIKS}.
(This phenomenon was also discovered in ~\cite{JMMS,JMU}.)

One of the main problems
in this approach is to calculate the expressions of the form:
\begin{equation}
<0|C(\mu_1)\cdots C(\mu_N) B(\lambda_N)\cdots B(\lambda_1)|0>
\end{equation}

In this paper we, using the operation of comultiplication, write down
a kind of the operator-valued
generating function for expressions of such a type. In this way we get some
expression for this scalar product, from which the V.E.Korepin's representation
in terms of the dual fields follows by direct computation.
Note, that the idea to use the operation of comultiplication in the
calculation of correlators was used, for example, in ~\cite{Korepin84a}
in the definition of the two-site generalized model.

\section{Definitions.}

The $ABCD$ algebra is the algebra generated by the elements of the $2\times 2$-
matrix $T$, which is the monodromy matrix of the auxiliary linear problem,
\begin{equation}
T(\lambda)=\left[\begin{array}{cc}A(\lambda)&B(\lambda)\\C(\lambda)
&D(\lambda)\end{array}\right]
\end{equation}
with the relations
\begin{equation}
R(\lambda,\mu)\s{1}{T}(\lambda)\otimes\s{2}{T}(\mu)=
\s{2}{T}(\lambda)\otimes\s{1}{T}(\mu)R(\lambda,\mu)
\end{equation}
where indices over the letters denote their order as operators ( from the left
to the right ) , and
\begin{equation}
R(\lambda,\mu)=\left[\begin{array}{cccc}f(\mu,\lambda)&0&0&0\\
0&1&g(\mu,\lambda)&0\\
0&g(\mu, \lambda)&1&0\\
0&0&0&f(\mu,\lambda)\end{array}
\right]
\end{equation}
\begin{equation}
f(\lambda,\mu)=\frac{\sinh (\lambda-\mu+2i\eta)}{\sinh (\lambda-\mu)}
\end{equation}
\begin{equation}
g(\lambda,\mu)=\frac{\sinh (2i\eta)}{\sinh (\lambda-\mu)}
\end{equation}

We will often write indices instead of arguments, for the brevity.
For example:
\begin{equation}
\begin{array}{cc}
A\l{} A\m{} =A\m{} A\l{},\/\/\/\/\/&B\l{} B\m{} =B\m{} B\l{}\\
C\l{} C\m{} =C\m{} C\l{},\/\/\/\/\/&D\l{} D\m{} =D\m{} D\l{}
\end{array}
\end{equation}
\begin{eqnarray}
A\m{} B\l{} =f\ml{}{} B\l{} A\m{} +g\lm{}{} B\m{} A\l{} ,\hspace{15pt}
(A\leftrightarrow B)\\
D\m{} B\l{} =f\lm{}{} B\l{} D\m{} +g\ml{}{} B\m{} D\l{} ,\hspace{15pt}
(D\leftrightarrow B)\\
C\l{} A\m{} =f\ml{}{} A\m{} C\l{} +g\lm{}{} A\l{} C\m{} ,\hspace{15pt}
(C\leftrightarrow A)\\
C\l{}D\m{}=f\lm{}{}D\m{}C\l{}+g\ml{}{}D\l{}C\m{},\hspace{15pt}
(C\leftrightarrow D)
\end{eqnarray}
\begin{equation}
\begin{array}{lll}
C\l{C}B\l{B}&=&g\dval{C}{B}\{A\l{C}D\l{B}-A\l{B}D\l{C}\}= \\
            &=&g\dval{C}{B}\{D\l{B}A\l{C}-D\l{C}A\l{B}\}
\end{array}
\end{equation}

This algebra can be considered as the Hopf algebra, if we provide it with the
coproduct
\begin{equation}
\Delta:T(\lambda)\rightarrow T(\lambda)\s{.}{\otimes} T(\lambda)
\end{equation}
which is the homomorphism from the $ABCD$-algebra to its tensor square:
\begin{eqnarray*}
\s{1}{(T(\lambda)\s{.}{\otimes}T(\lambda))}\otimes
\s{2}{(T(\mu)\s{.}{\otimes}T(\mu))} R(\lambda,\mu)=\\=
(\s{1}{T(\lambda)}\otimes\s{2}{T(\mu)})\s{.}{\otimes}
(\s{1}{T(\lambda)}\otimes\s{2}{T(\mu)})R(\lambda,\mu)=\\=
R(\lambda,\mu)(\s{2}{T(\lambda)}\otimes\s{1}{T(\mu)})\s{.}{\otimes}
(\s{2}{T(\lambda)}\otimes\s{1}{T(\mu)})=\\=
R(\lambda,\mu)\s{2}{(T(\lambda)\s{.}{\otimes}T(\lambda))}\otimes
\s{1}{(T(\mu)\s{.}{\otimes}T(\mu))}
\end{eqnarray*}

 This comultiplication has the remarkable property that the subalgebra
generated by $A$ and $C$ is the coideal~\cite{RTF}:
\begin{eqnarray}
\Delta(A)=A\otimes A + B\otimes C\\
\Delta(C)=C\otimes A + D\otimes C
\end{eqnarray}

This means that when we take the coproduct of the element of this subalgebra,
the second tensor multiplier will always belong to this subalgebra, i.e. be
some  product of $A$'s and $C$'s. It will be convenient for us to denote
\begin{eqnarray}
\psi^+(\lambda)=A(\lambda)\\
\psi^-(\lambda)=C(\lambda)
\end{eqnarray}

The commutation relations in the $\psi^{\pm}$-algebra are:
\begin{equation}
\psi\l{}^-\psi\m{}^+=f\ml{}{}\psi\m{}^+\psi\l{}^-+g\lm{}{}\psi\l{}^+\psi\m{}^-
\end{equation}
and the same with $+$ and $-$ exchanged.

\section{Scalar products of the wave functions.}

Consider the expression:
\begin{equation}
S(\{\mu\},\{\lambda\})=<0|C(\mu_1)\cdots C(\mu_N) B(\lambda_N)\cdots
B(\lambda_1)|0>
\end{equation}

Also consider the expression:
\begin{equation}
V(\{\mu\},\{\lambda\})=<0|\delta(\psi^-(\mu_1)\cdots\psi^-(\mu_N)
\psi^+(\lambda_N)\cdots\psi^+(\lambda_1))|0>
\end{equation}
where $|0>$ is the usual $ABCD$-vacuum, all $\psi$'s act freely both on the
left and on the right (so $V$ can be considered as an element of the
algebra generated by $\psi$-operators.) Note that expressions
\begin{equation}\nonumber
\psi\ep{1}\m{1}\cdots \psi\ep{N}\m{N}\psi\barep{N}\l{N}\cdots\psi\barep{1}\l{1}
\end{equation}
are linear independent for different sets $(\epsilon)$ and
$(\bar{\epsilon})$, and constitute the basis in the space of $2N$-linear
combinatins of $\psi$-operators with momenta $\{\mu_1,\ldots,\mu_N,
\lambda_1,\ldots,\lambda_N\}$ (this is because we can rearrange momenta
using the commutation relations to get expression
with $\mu$'s going the first.)
Note that $S(\mu,\lambda)$ is the coefficient of
\begin{equation}\nonumber
\psi^+\m{1}\cdots\psi^+\m{N}\psi^-\l{N}\cdots\psi^-\l{1}
\end{equation}
in $V(\mu,\lambda)$. So, the problem is to find this term. The simplest way
to do it is first to rearrange $\psi$'s in $V(\mu,\lambda)$ to get $\psi^+$'es
standing on the left of $\psi^-$'es --- because
\begin{equation}
\begin{array}{c}
<0|\delta(\psi^+\k{1}\cdots\psi^+\k{N}\psi^-\n{N}\cdots\psi^-\n{1})|0>=\\=
a(\kappa_1)\cdots a(\kappa_N)d(\nu_N)\cdots d(\nu_1)
\psi^+\k{1}\cdots\psi^+\k{N}\psi^-\n{N}\cdots\psi^-\n{1}
\end{array}
\end{equation}

So, if
\begin{equation}
\psi^-\m{1}\cdots\psi^-\m{N}\psi^+\l{N}\cdots\psi^+\l{1}=
\sum_{\{\kappa\},\{\nu\}}
K\left(\begin{array}{cc}
-&+\\
\{\kappa\}&\{\nu\}
\end{array}\right)
\psi^+\k{1}\cdots\psi^+\k{N}\psi^-\n{N}\cdots\psi^-\n{1}
\end{equation}
and
\begin{equation}
\psi^+\k{1}\cdots\psi^+\k{N}\psi^-_{\nu_N}\cdots\psi^-_{\nu_1}=
\sum_{(\epsilon^l),(\epsilon^r)}
\tilde{K}\left(\begin{array}{cc}
(\kappa)&(\nu)\\
(\epsilon^l)&(\epsilon^r)
\end{array}\right)
\psi^{\epsilon^l_1}\m{1}\cdots\psi^{\epsilon^l_N}\m{N}
\psi^{\epsilon^r_N}\l{N}\cdots\psi^{\epsilon^r_1}\l{1}
\end{equation}
then
\begin{equation}
\begin{array}{c}
<0|C(\mu_1)\cdots C(\mu_N)B(\lambda_N)\cdots B(\lambda_1)|0>=\\
=\sum_{\{\kappa\},\{\nu\}}
K\left(\begin{array}{cc}
-&+\\
\{\kappa\}&\{\nu\}
\end{array}\right)
\tilde{K}\left(\begin{array}{cc}
\{\kappa\}&\{\nu\}\\
+&-
\end{array}\right)
\prod_{\{\kappa\}}a(\kappa)\prod_{\{\nu\}}a(\nu)
\end{array}
\end{equation}

\section{Calculation of $K$ and $\tilde{K}$.}
Without any loss of the generality we can put:
\begin{eqnarray*}
\{\kappa\}=\{\lambda_n,\ldots,\lambda_1,\mu_{n+1},\ldots,\mu_N\}\\
\{\nu\}=\{\lambda_{n+1},\ldots,\lambda_N,\mu_n,\ldots,\mu_1\}
\end{eqnarray*}

All other possibilities can be reduced to this by the permutation
of $\mu$ and $\lambda$.
So, what we are to do is to take
\begin{equation}
\psi^-\m{1}\cdots\psi^-\m{n}\psi^-\m{n+1}\cdots\psi^-_{\mu_N}
\psi^+\l{N}\cdots\psi^+\l{n+1}\psi^+\l{n}\cdots\psi^+\l{1}
\end{equation}
and move $\psi^+$'es to the left of $\psi^-$'es using the commutation
relations,
and find the coefficient of the term
\begin{equation}
\psi^+\l{n}\cdots\psi^+\l{1}\psi^+\m{n+1}\cdots\psi^+\m{N}
\psi^-\m{1}\cdots\psi^-\m{n}\psi^-\l{N}\cdots\psi^-\l{n+1}
\end{equation}

Let us first move $\psi^+$'es from $\psi^+\l{N}\cdots\psi^+\l{n+1}$ to the very
left. We should not get any of $\{\lambda_N,\ldots,\lambda_{n+1}\}$
on the left as the indices of
$\psi^+$'es, so all of these $\lambda$'s should be exchanged
with some $\mu$'s and, moreover, with some $\mu$'s from $\{\mu_{n+1},
\ldots,\mu_n\}$, because we also don't need any of $\{\mu_1,\ldots,\mu_n\}$
on the left.

Let us, for example, do this procedure with $\psi^+\l{n+1}$. Moving its plus
to the left, we get a plenty of terms, of which we are interested in the
following ones:
\begin{equation}\begin{array}{c}
\sum_{k=1}^{N-n} C_k\psi^+\m{n+k}\psi^-\m{1}\cdots\psi^-\m{n}
\psi^-\m{n+1}\cdots \widehat{\psi^-\m{n+k}}\cdots \psi^-\m{N}\times \\ \times
\psi^-\l{n+1}
\psi^+\l{N}\cdots\psi^+\l{n+2}\psi^+\l{n}\cdots \psi^+\l{1}
\end{array}
\end{equation}
where $C_k$ is some coefficient. To find it, before the calculation rearrange
$\psi^-$'es in order to shift $\psi^-\m{n+k}$ to the most right among them.
Then
the only possibility to get the required term is to exchange momenta when
$\psi^+\l{n+1}$ passes through $\psi^-\m{n+k}$ (we will get the factor
$g\ml{n+k}{n+1}$) and then to pass created in such a manner
$\psi^+\m{n+k}$ to the very left without exchange, getting the product
of factors $f\dvam{n+k}{n+p}$ over $p\neq k$
and $f\dvam{n+k}{p}$ over $p\leq n$. Thus,
\begin{equation}
C_k=g\ml{n+k}{n+1}
\prod_{p\neq k} f\dvam{n+k}{n+p} \prod_{p\leq n} f\dvam{n+k}{p}
\end{equation}
 Repeating this procedure with $\psi^+\l{n+2}$  etc., we get
\begin{eqnarray*}
\sum_{\sigma\in{\cal S}_{N-n}}\hspace{-2pt}
g\ml{\sigma(1)+n}{1+n}\cdots g\ml{\sigma(N-n)+n}{N}
\prod_{i<j}f\dvam{\sigma(i)+n}{\sigma(j)+n}
\prod_{i<j}f\ml{\sigma(j)+n}{i+n}\times\\ \times
\prod_{p,q}f\dvam{n+p}{q}
\psi_{\mu_{n+1}}^+ \cdots\psi^+\m{N} \psi^-\m{1} \cdots\psi^-\m{n}
\psi^-_{\lambda_N} \cdots\psi^-\l{n+1} \psi^+\l{n} \cdots\psi^+\l{1}
\end{eqnarray*}

It is easy now to move $\psi^+\l{n}\cdots\psi^+\l{1}$ from the very right to
the
left. Ultimately we obtain for $K$ :
\begin{equation}
\begin{array}{c}
K\left(\begin{array}{cc}-&+\\ \{\kappa\}&\{\nu\}\end{array}\right)=
\prod f(\mu^{\kappa},\mu^{\nu}) \prod f(\lambda^{\kappa},\lambda^{\nu})
\prod f(\lambda^{\kappa},\mu^{\nu})\times \\ \times
\sum_{\sigma\in{\cal S}(\{\mu\}\cap\{\kappa\})}
g(\mu_{\sigma(1)}^{\kappa},\lambda_1^{\nu})\cdots
g(\mu_{\sigma(N-n)}^{\kappa},\lambda_{N-n}^{\nu})\times \\ \times
\prod_{i<j}f(\mu_{\sigma(i)}^{\kappa},\mu_{\sigma(j)}^{\kappa})
\prod_{i<j}f(\mu_{\sigma(j)}^{\kappa},\lambda_i^{\nu})
\end{array}
\end{equation}

Here we denoted, {\em e.g.,} $\lambda^{\kappa}$ the intersection of the sets
$\lambda$ and $\kappa$. To get $\tilde{K}$, we need to move all $\mu$'s in the
expression \[ \psi^+\lk{n}\cdots\psi^+\lk{1}\psi^-\mn{1}\cdots\psi^-\mn{n} \]
to the left and find the coefficient of the term in which all pluses
are on the left. Just in the same way as before, we get
\begin{equation}
\begin{array}{c}
\tilde{K}\left(\begin{array}{cc}
\{\kappa\}&\{\nu\}\\
+&-
\end{array}\right)=\frac{1}{\prod_{p,q} f(\lambda_p^{\kappa},\mu_q^{\nu})}
\sum_{\sigma\in{\cal S}(\lambda\cap\kappa)}
\prod_{i<j}\left[ f\lk{\sigma(j)}{_,}\lk{\sigma(i)}
f\lk{\sigma(i)}{_,}\mk{j} \right] \times\\ \times
g\lk{\sigma(1)}{_,}\mn{1}\cdots g\lk{\sigma(n)}{_,}\mn{n}
\end{array}
\end{equation}

 Consequently,
 \begin{equation}
 \begin{array}{c}
 <0|C(\mu_1)\cdots C(\mu_N) B(\lambda_N)\cdots B(\lambda_1)|0>=\\=
 \sum_{\kappa\cup\nu=\lambda\cup\mu}\prod_i a(\lambda_i^{\kappa})
 \prod_i a(\mu_i^{\kappa}) \prod_i d(\lambda_i^{\nu}) \prod_i d(\mu_i^{\nu})
\times \\ \times
\prod_{p,q}f\mk{p}{_,}\mn{q} f\lk{q}{_,}\lnnew{p}
\Phi_{N-n}(\lambda^{\nu},\mu^{\kappa})\Phi_n (\mu^{\nu},\lambda^{\kappa})
\end{array}
\end{equation}
where $n=\mbox{card}(\mu\cap\nu)=\mbox{card}(\lambda\cap\kappa)$ and
\begin{equation}
\begin{array}{c}
\Phi_k(\lambda,\mu)=\sum_{\sigma\in{\cal S}_k}
g\ml{\sigma(1)}{1}\cdots g\ml{\sigma(k)}{k}\times\\ \times
\prod_{i<j} f_{\mu_{\sigma(i)},\mu_{\sigma(j)}}\prod_{i<j}f\ml{\sigma(j)}{i}
\end{array}
\end{equation}

Actually
\begin{equation}
\begin{array}{c}
\Phi_k(\lambda,\mu)=\sum_{\sigma\in{\cal S}_k}
g\ml{1}{\sigma(1)}\cdots g\ml{k}{\sigma(k)}\times\\ \times
\prod_{i<j} f\dval{\sigma(i)}{\sigma(j)}\prod_{i<j}f\ml{j}{\sigma(i)}
\end{array}
\end{equation}

and so it is evident that $\Phi_k(\lambda,\mu)$ is symmetric in $\lambda$.

Tending $\mu_1$ to $\lambda_1$, we get
\begin{equation}
\begin{array}{c}
[\Phi_k(\lambda,\mu)]_{\lambda_1\rightarrow \mu_1}=
\frac{\sinh(2i\eta)}{\mu_1-\lambda_1}
\prod_{j\neq 1}\left[ f(\mu_1,\lambda_j) f(\mu_j,\lambda_1) \right]\times\\
\times
\Phi_{k-1}(\{\lambda\}\backslash\{\lambda_1\},\{\mu\}\backslash\{\mu_1\})
\end{array}
\end{equation}
and this agrees with the formula [6.16] of the paper \cite{Korepin82}

\section{Determinant representation and Dual Fields.}

Introduce $x_k=e^{2\mu_k}$, $y_k=e^{2\lambda_k}$, $z=e^{4i\eta}$.
Then

\begin{equation}\nonumber
\begin{array}{c}
\Phi_n=\frac{(\sqrt{z}-\frac{1}{\sqrt{z}})^n}{z^{\frac{n(n-1)}{2}}}
\sqrt{x_1\cdots x_n y_1\cdots y_n}\sum_{\sigma\in{\cal S}_n}
\frac{1}{(x\sig{1}-y_1)\cdots (x\sig{n}-y_n)}\times \\ \times
\prod_{i<j}\frac{zx\sig{i}-x\sig{j}}{x\sig{i}-x\sig{j}}
\prod_{i<j}\frac{zx\sig{j}-y_i}{x\sig{j}-y_i}
=\\=
\frac{(\sqrt{z}-\frac{1}{\sqrt{z}})^n}{(-z)^{\frac{n(n-1)}{2}}}
\frac{\sqrt{\prod_i x_i \prod_i y_i}}{\Delta(x)\Delta(y)}
\sum_{\sigma\in{\cal S}_n}(-1)^{l(\sigma)}
\prod_{i<j}[(zx\sig{i}-x\sig{j})(y_j-y_i)]\times\\ \times
\prod_{i<j}\frac{zx\sig{j}-y_i}{x\sig{j}-y_i}
\prod_k\frac{1}{x\sig{k}-y_k}
=\\=
\frac{(\sqrt{z}-\frac{1}{\sqrt{z}})^n}{(-z)^{\frac{n(n-1)}{2}}}
\frac{\sqrt{\prod_i x_i \prod_i y_i}}{\Delta(x)\Delta(y)}
\sum_{\sigma\in{\cal S}_n}(-1)^{l(\sigma)}
\prod_{i<j}\frac{zx\sig{j}-y_i}{x\sig{j}-y_i}
\prod_k\frac{1}{x\sig{k}-y_k}\times \\ \times
\prod_{i<j}[(x\sig{j}-y_i)(zx\sig{i}-y_j)+(zx\sig{i}-y_i)(y_j-x\sig{j})]
=\\=
\frac{(\sqrt{z}-\frac{1}{\sqrt{z}})^n}{(-z)^{\frac{n(n-1)}{2}}}
\frac{\sqrt{\prod_i x_i \prod_i y_i}}{\Delta(x)\Delta(y)}
\prod_{p,q}(zx_p-y_q)\times \\ \times
\left( \sum_{\sigma\in{\cal S}_n}(-1)^{l(\sigma)}
\prod_k\frac{1}{(zx\sig{k}-y_k)(x\sig{k}-y_k)}+R_n(\{x\},\{y\}) \right)
\end{array}
\end{equation}

where

\begin{equation}
\begin{array}{c}
R_n(\{x\},\{y\})=\sum_{m=1}^{\frac{n(n-1)}{2}}
\sum_{\{i_1<j_1,\ldots,i_m<j_m\}}
\sum_{\sigma\in{\cal S}_n}(-1)^{l(\sigma)}\times \\ \times
\prod_{r=1}^{m}\frac{(zx\sig{i_r}-y_{i_r})(y_{j_r}-x\sig{j_r})}
{(zx\sig{i_r}-y_{j_r})(x\sig{j_r}-y_{i_r})}
\prod_k\frac{1}{(x\sig{k}-y_k)(zx\sig{k}-y_k)}
\end{array}
\end{equation}

We intend to show that $R_n=0$. It is easy to check that $R_2$ is zero.
Let us proceed by induction. Suppose
$R_{n-1}=0$. As a function of $y_n$, $R_n$ has the first order pole
at $y_n\rightarrow x_k$, corresponding permutations having the property
$\sigma(n)=k$. Due to the presence of $(y_{j_r}-x\sig{j_r})$ in
the numerator, only the terms in $\sum_{\{(i_r<j_r)\}}$  with $j_r\neq n$
for any $r$ really have this pole. So, the residue in this pole is
equal to $ \frac{1}{(z-1)y_n}R_{n-1}(\{x\}\backslash\{x_k\},
\{y\}\backslash\{y_n\})$ and
thus, is zero.
Note that by construction, $R_n$ is anti-symmetric not only
with respect to $x$, but also with respect to $y$. Thus we can change
\[
\sum_{\{i_1<j_1,\ldots,i_m<j_m\}}\rightarrow
\sum_{\{i_1>j_1,\ldots,i_m>j_m\}}
\]

After this change one can prove, just in the same way, that the residue
in the pole at $y_n\rightarrow zx_k$ is also equal to zero. Thus, $R_n$, as
a function of $y_n$, has no poles and decrease at infinity. So it is zero.

Therefore, we have:
\begin{equation}
\begin{array}{c}
\Phi_n=
\frac{(\sqrt{z}-\frac{1}{\sqrt{z}})^n}{(-z)^{\frac{n(n-1)}{2}}}
\frac{\sqrt{\prod_i x_i \prod_i y_i}}{\Delta(x)\Delta(y)}
\prod_{p,q}(zx_p-y_q)\times \\ \times
\det_{i,j}\left|\left|\frac{1}{(zx_i-y_j)(x_i-y_j)}\right|\right|
=\\=
\prod_{p,q}h(\mu_p,\lambda_q)
\prod_{i<j}[ g(\mu_j,\mu_i)g(\lambda_i,\lambda_j) ]
\det_{i,j}\left|\left|\frac{g^2(\mu_i,\lambda_j)}{f(\mu_i,\lambda_j)}
\right|\right|
\end{array}
\end{equation}
where we introduce $h(\mu,\lambda)=f(\mu,\lambda)/g(\mu,\lambda)$.

Note that for $z=1$ we get in this way the well-known identity~\cite{B1855}:
\begin{equation}
\det\left|\left|\frac{1}{(x_i-y_j)^2}\right|\right|=
\mbox{per}\left|\left|\frac{1}{(x_i-y_j)}\right|\right|
\det\left|\left|\frac{1}{(x_i-y_j)}\right|\right|
\end{equation}

Finally we get for $Z$:
\begin{equation}
\begin{array}{c}
Z=\sum_{\kappa\cup\nu=\mu\cup\lambda} a(\lambda^{\kappa})a(\mu^{\kappa})
d(\lambda^{\nu}) d(\mu^{\nu})\times\\ \times
f(\mu^{\kappa},\mu^{\nu})f(\lambda^{\kappa},\lambda^{\nu})
h(\mu^{\kappa},\lambda^{\nu})h(\lambda^{\kappa},\mu^{\nu})\times \\ \times
\prod_{i<j}[g(\mu_j^{\kappa},\mu_i^{\kappa})g(\lambda_i^{\nu},\lambda_j^{\nu})]
\prod_{i<j}[g(\lambda_j^{\kappa},\lambda_i^{\kappa})
g(\mu_i^{\nu},\mu_j^{\nu})]\times \\ \times
\det_{i,j}\left|\left|\frac{g^2(\mu_i^{\kappa},\lambda_j^{\nu}}
{f(\mu_i^{\kappa},\lambda_j^{\nu})}\right|\right|
\det_{i,j}\left|\left|\frac{g^2(\lambda_i^{\kappa},\mu_j^{\nu})}
{f(\lambda_i^{\kappa},\mu_j^{\nu})}\right|\right|=
\end{array}
\end{equation}
\begin{equation}
\begin{array}{c}
=\prod_{i<j}[g(\mu_j,\mu_i)g(\lambda_i,\lambda_j)]
\sum_{\kappa\cup\nu=\mu\cup\lambda} a(\lambda^{\kappa})a(\mu^{\kappa})
d(\lambda^{\nu}) d(\mu^{\nu})\times\\ \times
\left[\begin{array}{c}{\mbox{sign}}\\{\mbox{factor}}\end{array}\right]
h(\kappa,\nu)
\det\left|\left|\frac{g^2(\mu^{\kappa},\lambda^{\nu})}
{f(\mu^{\kappa},\lambda^{\nu})}\right|\right|
\det\left|\left|\frac{g^2(\lambda^{\kappa},\mu^{\nu})}
{f(\lambda^{\kappa},\mu^{\nu})}\right|\right|
\end{array}
\end{equation}

The sign factor in this expression is the sign of permutation of
$\lambda$'s from their natural
order to the order $(\lambda^{\kappa},\lambda^{\nu})$, multiplied by the sign
of permutation of $\mu$'s to $(\mu^{\kappa},\mu^{\nu})$. Here, for example,
$a(\lambda^{\kappa})$ means $\prod_i a(\lambda_i^{\kappa})$.
One can easily prove ~\cite{Korepin87} that this expression is equal to
\begin{equation}
\prod_{i<j}[g(\mu_j,\mu_i)g(\lambda_i,\lambda_j)]
<0|\det_N S |0>
\end{equation}
where
\begin{equation}
S_{jk}=\frac{g^2(\mu_j,\lambda_k)}{f(\mu_j,\lambda_k)}
e^{\phi^{\kappa}(\mu_j)+\phi^{\nu}(\lambda_k)}+
\frac{g^2(\lambda_k,\mu_j)}{f(\lambda_k,\mu_j)}
e^{\phi^{\kappa}(\lambda_k)+\phi^{\nu}(\mu_j)}
\end{equation}
and $\phi^{\kappa}$, $\phi^{\nu}$ are the commuting dual fields:
\begin{equation}
\begin{array}{c}
\phi^{\kappa}(\lambda)=q^{\kappa}(\lambda)+\pi^{\nu}(\lambda)\\
\phi^{\nu}(\lambda)=q^{\nu}(\lambda)+\pi^{\kappa}(\lambda)\\
\left[ \pi^{\kappa}(\lambda),q^{\kappa}(\mu) \right] =\ln h(\mu,\lambda)\\
\left[ \pi^{\nu}(\lambda),q^{\nu}(\mu) \right]=\ln h(\lambda,\mu)
\end{array}
\end{equation}
and $<0|$, $|0>$ are the dual vacuums:
\begin{equation}
\begin{array}{ll}
\pi^{\kappa}(\lambda)|0>=0,&\pi^{\nu}(\lambda)|0>=0\\
<0|q^{\kappa}(\lambda)=\mbox{ln}a(\lambda)<0|,&
<0|q^{\nu}(\lambda)=\mbox{ln}d(\lambda)<0|
\end{array}
<0|0>=1
\end{equation}

\section{Acknowledgements.}
I want to thank A.Mironov, A.Morozov, A.Zabrodin and especially N.Slavnov
for discussions and explanations. I am very indepted to A.G.Izergin for
valuable remarks.

\end{document}